\documentclass[aps,preprintnumbers,showpacs,twocolumn]{revtex4-1}

\usepackage[dvips]{graphicx}
\usepackage{dcolumn}
\usepackage{bm}
\usepackage{amssymb}

\begin{document}

\preprint{}

\title{Inference of topology and the nature of synapses, and the flow of
information in neuronal networks}

\author{F. S. Borges$^1$, E. L. Lameu$^2$, K. C. Iarosz$^{1,3}$, P. R.
Protachevicz$^4$, I. L. Caldas$^1$, R. L. Viana$^5$, E. E. N. Macau$^2$, A. M.
Batista$^{1,3,4,5}$, M. S. Baptista$^3$}
\affiliation{$^1$Physics Institute, University of S\~ao Paulo, S\~ao Paulo, SP,
Brazil.\\
$^2$National Institute for Space Research, S\~ao Jos\'e dos Campos, SP,
Brazil.\\
$^3$Institute for Complex Systems and Mathematical Biology, University of
Aberdeen, SUPA, UK.\\
$^4$Post-Graduation in Science, State University of Ponta Gros\-sa, Ponta 
Grossa, PR, Brazil.\\
$^5$Physics Department, Federal University of Paran\'a, Curitiba, PR,
Brazil.\\
$^6$Mathematics and Statistics Department, State University of Ponta Grossa,
Ponta Grossa, PR, Brazil.
}

\date{\today}

\begin{abstract}
The characterisation of neuronal connectivity is one of the most important
matters in neuroscience. In this work, we show that a recently proposed
informational quantity, the causal mutual information, employed with an
appropriate methodology, can be used not only to correctly infer the direction
of the underlying physical synapses, but also to identify their excitatory or
inhibitory nature, considering easy to handle and measure bivariate
time-series. The success of our approach relies on a surprising property found
in neuronal networks by which non-adjacent neurons do ``understand'' each other
(positive mutual information), however this exchange of information is not
capable of causing effect (zero transfer entropy). Remarkably, inhibitory
connections, responsible for enhancing synchronisation, transfer more
information than excitatory connections, known to enhance entropy in the
network. We also demonstrate that our methodology can be used to correctly
infer directionality of synapses even in the presence of dynamic and
observational Gaussian noise, and is also successful in providing the effective
directionality of inter modular connectivity, when only mean fields can be
measured.
\end{abstract}


\maketitle


Many real systems have been modelled by complex networks with different
topological characteristics. Network theory has been applied in a large number
of examples and different research fields, such as biology \cite{barabasi04},
economics \cite{niemira04} and physics \cite{arenas08}. In neuroscience, the
application of network theory provides a way to analyse the structure and the
functional behaviour of neuronal systems \cite{bullmore09}. A fundamental
research topic in neuroscience is the determination of the structure of the
brain, to better understand its functioning. Some neuronal networks had their
structure directly mapped by means of diffusion tensor imaging
tractography \cite{gong09}.

One of the most challenging problems in neuronal networks is the inference of
its topology, that is, the determination of the underlying synaptic
connectivity by indirect means, based on functional measurements of time-series
of the membrane potential \cite{ta10,bastos16}. There are works that infer the
topology based on functional measures such as correlation
\cite{takigawa96,baccala01} and synchronisation \cite{cui15}, or functional
magnetic resonance imaging \cite{heuvel10}. And there are those based on
informational quantities
\cite{amblard11,schreiber00,runge15,rubido14,martinez16}. Inference based on
functional measures requires a threshold analysis that establishes a link
between the measurement and the physical connection
\cite{amblard11,massey90,vlachos10}. Rubido et al. \cite{rubido14} showed that
a threshold can be calculated whenever a functional measure between nodes
(cross correlation (CC) or mutual information) in a network is dissimilar.
Higher functional values correspond to a pair of adjacent nodes, lower
functional values to non-adjacent nodes. Bianco-Martinez et al.
\cite{martinez16} used the mutual information rate (MIR) to successfully infer
the connectivity of a network composed of Hindmarsh-Rose (HR) neurons
\cite{hindmarsh84} connected by electrical synapses. Both works in Refs.
\cite{rubido14,martinez16} have shown that the threshold technique could
surprisingly provide an inferred network that matched exactly with the real
network. These works have considered undirected networks, where nodes were
connected bidirectionally with the same intensity. 

This work considers HR networks with chemical sy\-napses. Unlike electrical
synapses that are undirected, chemical synapses are directed \cite{pereda14}.
Whereas undirected networks can have their topologies properly inferred by CC
and MIR, directed networks require methodologies capable of detecting the
directionality of the physical influence \cite{amblard11,massey90,liu12}.
Granger causality \cite{amblard11} is a concept construct on the idea that one
can obtain optimal fittings of mathematical models about the measured
time-series that provide the structure and direction of the connectivity. These
models are statistically optimised to improve the predictability of events in
one time-series based on observations of other time-series and have been shown
to be a powerful tool to infer \cite{bjorn14}. Informational quantities have
also been demonstrated to provide a framework that is at the frontier to infer.
In Ref. \cite{rubido14} it was shown that inference based on mutual information
is more reliable than those based on correlational measurements. In Ref.
\cite{liu12} it was shown that directed information had advantages over Granger
causality for quantifying effective connectivity in the brain. One question
that remains open is whether information measures can reliably infer the
connectivity of complex neuronal networks for all existing synapses by only
accessing bivariate measurements, in contrast to more complex and computational
demanding techniques such as multivariate analysis based on informational
analysis \cite{runge15,sun15}, a technique that takes into consideration
time-series from more than two neurons at each time, or modelled-based
multivariate approaches such as those that employ compressive sensing
\cite{wang16}.

In this work, we use the recently defined causal mutual information (CaMI)
\cite{martinez15,murilo16} calculated using an appropriate methodology to infer
the direction of chemical synapses in complex neuronal networks without any
mistake, by only considering easy to handle and to measure bivariate
time-series. Moreover, we show that inhibitory connections are responsible for
a considerably larger amount of information transfer than that compared to
neurons connected by excitatory synapses. This allows one to infer also the
nature of the connection (excitatory and inhibitory), and not only its
existence as previous techniques. Furthermore, we will also show that
non-adjacent neurons transmit roughly null amount of directed information,
indicating that indeed causal information has a direct relationship with the
existence of a synapse.

The CaMI was constructed from the idea that if there is a flow of information
from a system A to a system B, then longer time-series (or measurements with
higher precision) from B should have a positive mutual information to short
time-series (or to observations with lower precision) in A. This quantity,
measuring the influence from A to B, was shown to be equal to the transfer
entropy (TE) \cite{schreiber00} from A to B plus the mutual information between
A and B when both systems are being measured with the same resolution. The
advantage of CaMI however is that it allows one to calculate TE, and therefore
the directionality of the flow of information, by using measurements with
arbitrary resolution. Which in turn also allows for the correct calculation of
the TE using binary partitions of the phase space, i.e., appropriated when
measurements have the lowest possible resolution. Moreover, CaMI can be
calculated in lower-dimensional space of only $2$ dimensions, without the need
to consider conditional probabilities, but only marginal and joint
probabilities, and finally, it is a quantity that fully express not only the
exchange of information (MIR), but also its causal directionality (TE).

We consider the random neuronal network (RNN) \cite{barabasi99,timotheou10}
introduced by Gelenbe \cite{gelenbe89} and the neuronal network of the nematode
worm C. elegans \cite{kaiser06} whose structure was completely mapped at a
cellular level \cite{white86}. The node dynamics in the network is expressed by
the Hidmarsh-Rose (HR) neuron model. Hindmarsh and Rose \cite{hindmarsh84}
proposed a phenomenological neuron model that is a simplification of the
Hodgkin-Huxley model \cite{hodgkin52}. The HR is described by
$\dot{p}=q-ap^3+bp^2-n+I_{\rm ext}$, $\dot{q}=c-dp^2-q$, $\dot{n}=r[s(p-p_0)-n]$,
where $p(t)$ is the action potential of the membrane, $q(t)$ is related to the
fast current, $K^+$ or $Na^+$, and $n(t)$ is associated with the slow current,
for instance, $Ca^{2+}$. We use the parameters $a=1$, $b=3$, $c=1$, $d=5$,
$s=4$, $r=0.005$, $p_0=-1.60$ and $3.24\leq I_{\rm ext} \leq 3.25$, so that the
HR neuron exhibits a chaotic burst behaviour. Pre-synaptic neurons with an
action potential $p_j$ coupled by chemical synapses to neurons $i$ modifying
its action potential $p_i$ according to
${\dot{p}}_i=q_i-ap_i^3+bp_i^2-n_i+I_{\rm ext}+g_c(V_{\rm syn}-p_i)
\sum_{j=1}^{N}\varepsilon_{ij}\Gamma(p_j)$, where $(i,j)=1,\cdots,N$ and $N$ is
the neurons number. The chemical synapse function is modelled by the sigmoidal
function $\Gamma(p_j)=\frac{1}{1+\exp[-\lambda(p_j-{\Theta}_{\rm syn})]}$, with
${\Theta}_{\rm syn}= 1.0$, $\lambda = 10$, $V_{\rm syn} = 2.0$ for excitatory
and $ V_{\rm syn} = -1.5$ for inhibitory synapses. The adjacency matrix
$\varepsilon_{ij}$ describes the neurons chemically connected. To do our
analysis, we normalise $p_i$ through the equation
$x_i=\frac{p_i^{\rm max}-p_i}{p_i^{\rm max}-p_i^{\rm min}}$, where $p_i^{\rm max}$ and
$p_i^{\rm min}$ are the maximum and minimum values, respectively, of the
time-series of $p_i(t)$.

\begin{figure}[htbp]
\begin{center}
\includegraphics[height=5.5cm,width=7cm]{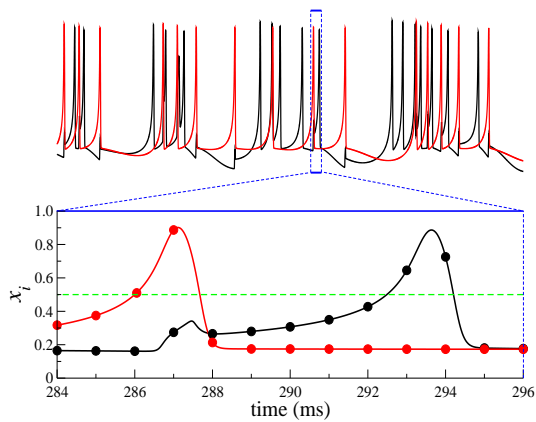}
\caption{(Colour online) Normalised membrane potential of two chemical coupled
HR neurons with connection from $x_2$ (red line) to $x_1$ (black line). We
consider the coupling strength $g_c=1$ and the mapping time step
$\Delta t=1$ms. The black and red circles correspond to $X_1$ and $X_2$,
respectively.}
\label{fig1}
\end{center}
\end{figure}

In order to be able to describe most of the information content of the
time-series by a short-length binary symbolic representation, we make a
time-Poincar\'e map of the time-series. Ideally, in the case one wants short
symbolic sequences to fully express the amount of information of infinitely
long sequences, points in the mapping should be spaced by a time step such that
the symbolic representation of the time-series behaves as a random process,
i.e., the next symbolic sequence is decorrelated with the previous. We are
interested in obtaining a good estimation of CaMI to correctly infer the
network's topology, its synaptic nature and to obtain a sufficiently accurate
value for the magnitude of the flow of information (e.g., CaMI, MIR and TE).
Given a time step $\Delta t$, a mapping for neuron $i$ $X_i$ is constructed by
collecting a point of the membrane potential at times $t=n\Delta t$ producing
the discrete time-series described $x^n_i=x_i$ ($t=n\Delta t$). In this way, we
obtain the mapping $X_i = x_i^0, x_i^1, x_i^2, ...,x_i^{T-1}$ for neuron $i$,
where $T$ is the number of points in the mapping. In the following, we will
study coupled neurons to determine a time step for which CaMI is maximised,
aiming with this maximisation to construct a time Poincar\'e map that tends to
behave as a Markov process, allowing CaMI, MI and TE to express a good
approximant of their real values. Figure \ref{fig1} shows the normalised
membrane potential for two chemical coupled HR neurons with connection from
$x_2$ (red line) to $x_1$ (black line). The black and red circles correspond to
$X_1$ and $X_2$, respectively, where the mapping step time $\Delta t$ is equal
to $1$ms. With the forward-time trajectory $X_i^L(n)={x_i^n,\cdots,x_i^{n+L-1}}$,
where $L$ is the length of the time-series $X(n)$ and $n$ is the discrete time,
we generate a symbolic sequence $S_i^L(n)={s_i^n,\cdots,s_i^{n+L-1}}$
(Supplementary Material), where we consider $s_i^n=0$ if $x_i^n\leq 0.5$ and
$s_i^n=1$ if $x_i^n>0$.

Bianco-Martinez and Baptista \cite{martinez15,murilo16} defined a new quantity
named CaMI from $X_i$ to $X_j$ (${\rm CaMI}_{X_i\rightarrow X_j}$) as the MI
between joint events in $X_i^{-L}$ and the set composed by the joint events of
$X_j^{-L}$ and $X_j^L$ as ${\rm CaMI}_{X_i\rightarrow  X_j}={\rm MI}
(X_i^{-L};(X_j^{-L},X_j^L))={\rm MI}(X_i^{−L};W_j^{2L})$, where the mutu\-al
Information ${\rm MI} (X_i^L;X_j^L)$ is given by ${\rm MI}(X_i^L;X_j^L)=
H(X_i^L)+H(X_j^L)-H(X_i^L,X_j^L)$, and $H(X_i^L)$ is the Shannon entropy of
length-$L$ trajectory points of the discrete mapping. It is also true that
${\rm CaMI}_{X_i \rightarrow X_j}={\rm MI}(X_i^L;X_jL)+{\rm TE}_{X_i \rightarrow X_j}$.
Probabilities to calculate CaMI are constructed considering the probabilities
of the encoded binary symbolic sequences. CaMI is thus calculated by
${\rm CaMI}_{X_i \rightarrow X_j}=\sum_{S_i}\sum_{S_j}P(S_i^L,S_j^{2 L})\log
\frac{P(S_i^L,S_j^{2 L})}{P(S_i^L) P(S_j^{2 L})}$, where the summation indexes
$S_i$ and $S_j$ represent the space of possible length-$L$ symbolic sequences
coming from neuron $i$ and $S_j$ the space of possible joint events of finding
a length-$L$ symbolic sequence coming from neuron $j$ at time $n-L$ followed by
a length-$L$ symbolic sequence in this same neuron at time $n$, or in other
words, of finding a length-$2L$ symbolic sequence in neuron $j$ starting at
the time $n-L$. $P(S_i^L)$ is the probability of finding symbolic sequences
$S_i^L=\{s_i,\ldots,s_i^{L-1}\}$ in $X_i$, $P(S_j^{2 L})$ is the probability of
finding a particular length-$L$ symbolic sequences
$S_i^{2L}=\{s_i,\ldots,s_i^{2L-1}\}$ in $X_j$, and $P(S_i^L,S_j^{2 L})$ is the
joint probability between length-L symbolic sequences in neuron $i$ and
length-$2L$ symbolic sequences in neuron $j$. The directionality index defined
in Ref. \cite{schreiber00} in terms of the TE can be calculated by
${\rm DI}_{X_i \rightarrow  X_j}={\rm CaMI}_{X_i \rightarrow  X_j}-
{\rm CaMI}_{X_j \rightarrow  X_i}$. For simplicity in notation we consider that
${\rm DI}_{X_i \rightarrow X_j} \equiv {\rm DI}_{ij}$. This index measures the net
amount of directed information flowing from $X_i$ to $X_j$. Thus, if
${\rm DI}_{ij}$ is positive (negative), there is a net amount of information
flowing from neuron $i$ to neuron $j$ (from neuron $j$ to neuron $i$). Our
hypothesis, also sustained by the works of \cite{rubido14,martinez16} and
others is that if there is a directed adjacent connection from neuron $i$ to
$j$, thus ${\rm DI}_{ij}$ will be considerably larger than the directionality
index of neurons that are not adjacently connected. So, the connection is
$X_i \rightarrow X_j$ if ${\rm DI}_{ij}>h$, the connection is
$X_j \rightarrow X_i$ if ${\rm DI}_{ij}<-h$, and there is no connection if
${\rm DI}_{ij}\cong 0$. In the latter case, the directionality index will be
close to zero because the transfer entropy will be roughly zero for
non-adjacent nodes. The mutual information is a symmetric quantity and
therefore ${\rm MI}(X_j,X_i)={\rm MI}(X_i,X_j)$.

\begin{figure}[htbp]
\begin{center}
\includegraphics[height=3cm,width=7.5cm]{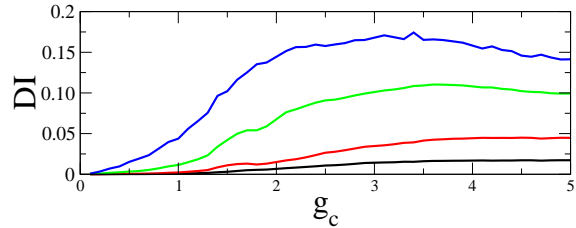}
\caption{(Colour online) Directional index ($DI$) as a function of the coupling
strength ($g_c$) for the mapping step time $\Delta t=0.25$ms. We consider
$L=1$ (black line), $L=2$ (red line), $L=4$ (green line), and $L=8$ (blue
line).}
\label{fig2}
\end{center}
\end{figure}

In Fig. \ref{fig2} we calculate DI as a function of $g_c$ for two coupled
neurons with one directional connection from $x_1$ to $x_2$. We observe that
${\rm DI}=0$ when the neurons are uncoupled ($g_c=0$), and ${\rm DI}>0$ for
$g_c>0$. The information is transmitted from $x_1$ to $x_2$, in accordance with
the direction of the connection. We compute DI for $L=1$ (black line), $L=2$
(red line), and $L=8$ (blue line). For the following analysis, we fix
$\Delta t=0.25$ms, and $L=8$ that maximises DI values.

\begin{figure}[htbp]
\begin{center}
\includegraphics[height=8.0cm,width=7.5cm]{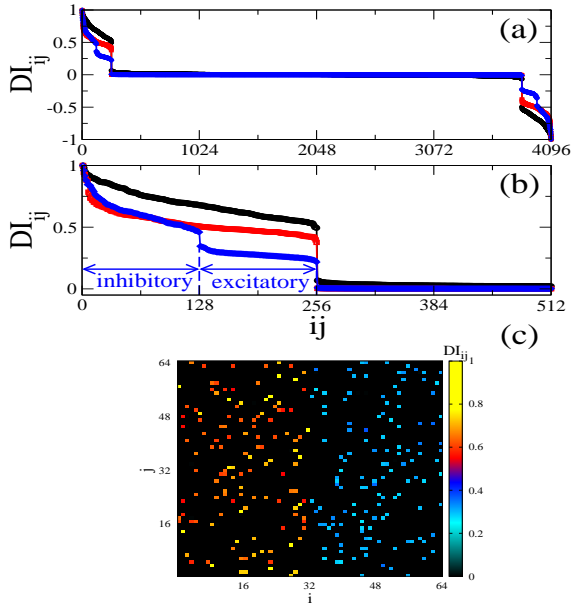}
\caption{(Colour online) (a) Normalised directional index, ranked from larger
to smaller values, for a random neuronal network with $N=64$ HR neurons, $K=4$,
$\Delta t=0.25$, $L=8$, $T=4\;10^6$, and $g_c=0.1$. We consider $3$ cases for
the connectivity: $256$ excitatory synapses (black line), $256$ inhibitory
synapses (red line), and $128$ excitatory and $128$ inhibitory synapses (blue
line). (b) Magnification of (a). (c) Matrix of the normalised directional index
(${\rm DI}_{ij}$) of latter case.}
\label{fig3}
\end{center}
\end{figure}

Next, we build a directed network where the connections among the neurons are
randomly chosen. We consider a random neuronal network with $64$ HR neurons and
average degree of connectivities $K$ equal to $4$. As a consequence, the
network has $256$ of a total of $4096$ directed connections ($ij$). Figure
\ref{fig3} shows the normalised directional index, ranked from larger to
smaller values, for $3$ different neuronal connectivity configurations: $256$
excitatory synapses (black line), $256$ inhibitory synapses (red line), and
$128$ excitatory and $128$ inhibitory synapses (blue line). In Fig.
\ref{fig3}(a) there are $2$ regions with ${\rm DI}_{ij}\neq 0$, that represent
the connections from $i$ to $j$, while ${\rm DI}_{ij}\approx 0$ corresponds to
the situation in that there is no connection between $i$ and $j$. The
magnification (Fig. \ref{fig3}(b)) exhibits two abrupt transitions. The
transition to ${\rm DI}_{ij}\approx h$ allows the detection of directed
connections in the neuronal network. The transition that occurs for
${\rm DI}_{ij}>h$ allows to infer the excitatory and inhibitory synapses, as
shown by the blue line, where we observe the existence of $128$ excitatory and
$128$ inhibitory synapses.

Notice that the DI values between adjacent and non-adjacent neurons are notably
dissimilar, meaning that a small threshold $h$ can be chosen such that
$DI_{ij}>h$ implies a directed connection from neuron $i$ to neuron $j$. For the
network whose neurons are connected by both inhibitory and excitatory synapses,
we notice in the blue line of Fig. \ref{fig3} two ranges of DI dissimilar
values. For $h<$DI$_{ij}<0.4$, the connection is excitatory and for DI$_{ij}>0.4$
the connection is inhibitory. In Fig. \ref{fig3}(c) we see the adjacency matrix,
where the coloured elements of the matrix indicate if the pairs of neurons are
connected. The uncoupled pairs of neurons are indicated in black, while the
coupled pairs are in colour scale according to the normalised directional
index. We consider the same parameters used to calculate the blue line in Fig.
\ref{fig3}. For ${\rm DI}_{ij}< 0.4$ the colour scale shows the excitatory
synapses and for ${\rm DI}_{ij}\geq 0.4$ the synapses are inhibitory.

\begin{figure}[htbp]
\begin{center}
\includegraphics[height=3cm,width=7.5cm]{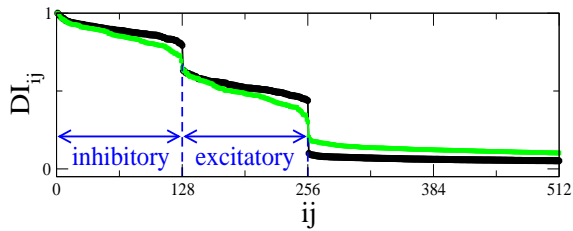}
\caption{(Colour online) Normalised directional index for
$\sigma_d=3$ (black line) and $\sigma_d=4$ (green line). We consider
$\Delta t=0.25$, $L=8$, and $g_c=0.15$.}
\label{fig4}
\end{center}
\end{figure}

We analyse the noise effect in the inference of the connections. Neuronal noise
can be related to several sources, such as synaptic noise \cite{gyorgyi90} and
ion conductance noise \cite{cao05}. In the action potential equation, we add a
Gaussian noise with zero mean and variance $\sigma_d$. We
calculate the ${\rm DI}_{ij}$ values for the neuronal network with $\sigma_d=3$
(black line) and $\sigma_d=4$ (green line), as shown in Fig. \ref{fig4}. We
verify that the inference for the existence of a synapse is robust to dynamic
noise in the membrane potential. However, for $\sigma_d\gtrsim 3.5$ it is not
possible to infer whether the synapse is excitatory or inhibitory. Therefore,
the inference of the connectivities is more robust than the inference of its
nature of the synapses. CaMI-based inference is also robust to additive noise
of moderate amplitude (Supplementary Material).

\begin{figure}[htbp]
\begin{center}
\includegraphics[height=3cm,width=7.5cm]{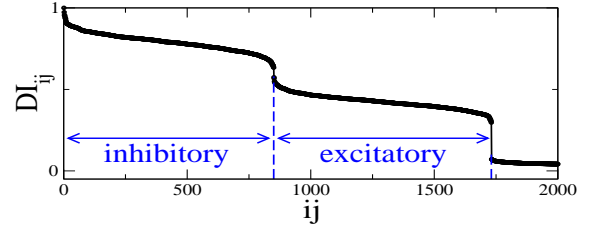}
\caption{ Normalised directional index, ordered from larger to smaller values
for $N = 277$ HR neurons. We consider  $\sigma_d=1$, $g_c=0.035$, $\Delta
t=0.25$, and $L=8$.}
\label{fig5}
\end{center}
\end{figure}

In the literature, there are many works that consider C. elegans neuronal
network to study nervous system \cite{liu95,varshney11}. The C. elegans is a
soil worm with body size about $1$mm and a simple nervous system
\cite{gally03}. We consider in our study the connectome of the large somatic
nervous system according to Ref. \cite{celegans} that consists of $277$
neurons. To test our inference approach, we consider approximately $50\%$ of 
excitatory and $50\%$ of inhibitory synapses in the C. elegans network with
$1731$ directed connections. The directed adjacency matrix ($\varepsilon_{ij}$)
is obtained from the brain connectivity of the C. elegans. Figure \ref{fig5}
exhibits the DI values, where the two discontinuities transitions in the DI
values correspond to the excitatory and inhibitory synapses. In Fig.
\ref{fig5} is possible to identify the connected neurons of the C. elegans,
where from $i=1$ to $i=138$ and from $i=139$ to $i=277$ there are $850$
inhibitory synapses and $881$ excitatory synapses, respectively. 

In conclusion, we propose a successful methodology based on CaMI to infer,
characterise and investigate the transmission of information in neuronal
networks with chemical synapses. Through the CaMI, we show not only how to
infer the existence of synapses, but also to identify the nature of the
synapse. Our technique can be applied to time-series generated with Gaussian
dynamical noise inbuilt in the neuron equations, or to time-series contaminated
by observational noise (Supplementary Material). Moreover, we also showed that
when access to the neuron potential is not possible, but rather only local mean
fields can be measured, such as those coming from EEG signals, our CaMI-based
technique can correctly determine the effective net directed connectivity
between different neuronal clusters. This work also shows that excitatory
connections are not so efficient to transfer information as inhibitory
connections, and that non-adjacent neurons transfer roughly zero amount of
information. This latter observation suggests that a pre-synaptic neuron (a
neuron that has an adjacent connection to the post-synaptic one) not only
exchange information (positive mutual information), but is also capable of
using information to cause an effect in a post-synaptic neuron (positive
transfer entropy). Non-adjacent neurons only exchange information. This
one-to-one relationship between structure and information transmission remains
valid for a wide range of the coupling strength $g_c$, constraint within an
interval with not so small (to prevent full decorrelation) and not so large
(to prevent full synchronization) bounds. For an inference with no mistakes,
time-series  should be sufficiently long, more specifically, their size in
seconds (i.e., $t$) should scale with the size of the network, and it is a
function of the coupling strength, a relationship that was studied in much
detail in Ref. \cite{rubido14}. 

Acknowledgement: CAPES, DFG-IRTG 1740/2, Fun\-da\c c\~ao Arauc\'aria, Newton
Fund, CNPq (154705/2016-0, 311467/2014-8), FAPESP (2011/19296-1, 2015/07311-7,
2016/16148-5, 2016/23398-8, 2015/50122-0), EPSRC-EP/I032606.


\newpage
\section*{Inference of topology and the nature of synapses, and the flow of information in
neuronal networks: supplementary material}

\section*{Encoding the trajectory into symbolic sequences}

Figure 1 (in the paper) shows the normalised membrane potential for two 
neurons coupled from $x_2$ (red line) to $x_1$ (black line). In table
\ref{tab1S}, we show some mapped values of $x_1^n$ and $x_2^n$ with their
respective length-1 symbolic values $s_1^n$ and $s_2^n$, and also the length-2
and length-4 symbolic sequence $S_1^{L=2} (n)$ and  $S_2^{L=4} (n)$,
respectively.

\begin{table}[hbt]
\centering
\begin{tabular}{|c|c|c|c|c|c|c|}
\hline
{\bf n  }&{\bf  $x_1^n$}&{\bf  $s_1^n$}&{\bf  $S_1^{L=2} (n)$}&
{\bf  $x_2^n$}&{\bf  $s_2^n$}&{\bf  $S_2^{L=4} (n)$} \\
\hline \hline
 285   &   0.163043   &   0   &   00   & 0.374431  &   0   &   0110  \\
 286   &   0.161350   &   0   &   00   & 0.500448  &   1   &   1100  \\
 287   &   0.274266   &   0   &   00   & 0.886694  &   1   &   1000  \\
 288   &   0.265589   &   0   &   00   & 0.213396  &   0   &   0000   \\
 289   &   0.279589   &   0   &   00   & 0.174788  &   0   &   0000  \\
 290   &   0.306991   &   0   &   00   & 0.174349  &   0   &   0000  \\
 291   &   0.349396   &   0   &   00   & 0.173966  &   0   &   0000   \\
 292   &   0.427650   &   0   &   01   & 0.173642  &   0   &   0000  \\
 293   &   0.645130   &   1   &   11   & 0.173384  &   0   &   ... \\
 294   &   0.725724   &   1   &   10   & 0.173200  &   0   &   ... \\
 295   &   0.180110   &   0   &  ...   & 0.173100  &   0   &   ...\\
\hline
\end{tabular}
\caption{\small Mappings for  $\Delta t=1$ms and $L=2$.}
\label{tab1S}
\end{table} 

Considering the symbolic sequences $S_i^L$ and $S_j^{2L}$ is possible to find
the probabilities $P(S_i^L)$, $P(S_j^{2L})$ and $P(S_i^L, S_j^{2L})$. These
probabilities are used to calculate the Casual Mutual Information 
\begin{eqnarray}
CaMI_{X_i \rightarrow X_j} =  \sum_{S_i}  \sum_{S_j} P(S_i^L,S_j^{2 L})
\log \frac{P(S_i^L,S_j^{2 L})}{P(S_i^L) P(S_j^{2 L})},
\label{CaMI0}
\end{eqnarray}
and Directionality Index 
\begin{eqnarray}
\label{DICaMI}
DI_{X_i \rightarrow  X_j} = CaMI_{X_i \rightarrow  X_j} - CaMI_{X_j \rightarrow  X_i}.
\end{eqnarray}

\section*{Informational quantities for adjacent neurons}

\begin{figure}[htbp]
\begin{center}
\includegraphics[height=5.5cm,width=7cm]{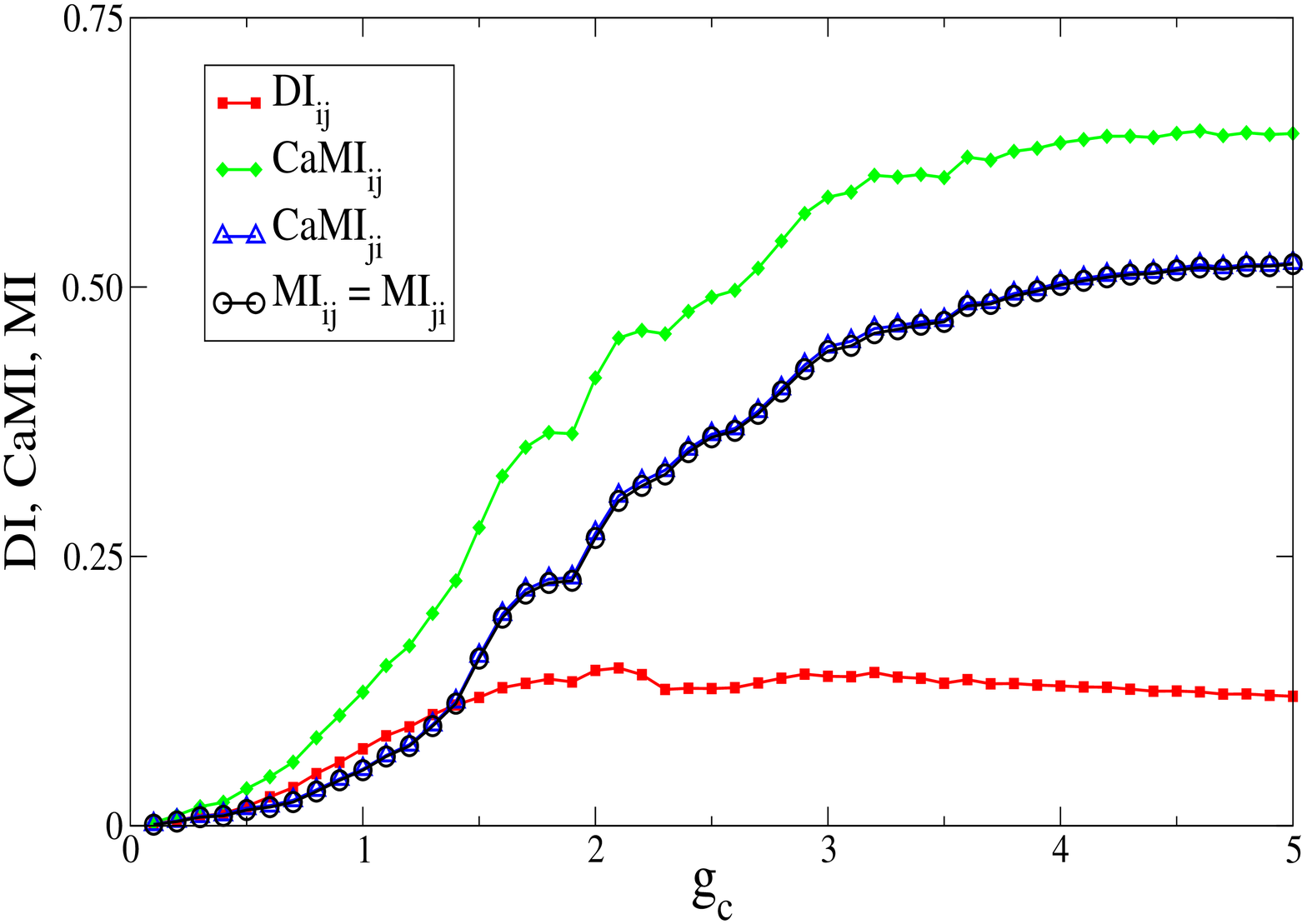}
\caption{Directional index (DI), causal mutual information (CaMI), and mutual
information (MI), as a function of the coupling strength ($g_c$) for the
mapping step time $\Delta t=0.5$ms and $L=8$.}
\label{fig1S}
\end{center}
\end{figure}

The ${\rm CaMI}_{ij}$, measuring the influence from $i$ to $j$, was shown to be
equal to ${\rm CaMI}_{ij} = {\rm MI}_{ij} + {\rm TE}_{ij}$, where
${\rm MI}_{ij}={\rm MI}_{ji}$ is the mutual information, and ${\rm TE}_{ij}$ is
the transfer entropy from $i$ to $j$. In Fig. \ref{fig1S}, we compare the
values of ${\rm DI}_{ij}$, ${\rm CaMI}_{ij}$, ${\rm CaMI}_{ji}$, and
${\rm MI}_{ij}$, as a function of $g_c$ for two coupled neurons with one
directional connection from $x_i$ to $x_j$. We observe that
${\rm CaMI}_{ji} \approx {\rm MI}_{ij}$, therefore ${\rm TE}_{ji} \approx 0$ and
${\rm TE}_{ij} \approx {\rm DI}_{ij}$.

\begin{figure}[htbp]
\begin{center}
\includegraphics[height=4cm,width=7cm]{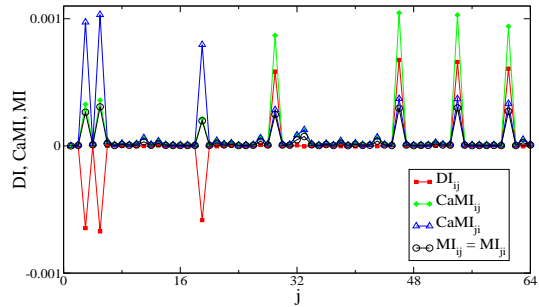}
\caption{Directional index (DI), causal mutual information (CaMI), and mutual
information (MI), for o neuron $i=1$ in a network with $N=64$ neurons, for the coupling strength $g_c = 0.1$,
mapping step time $\Delta t=0.5$ms, and $L=8$.}
\label{fig2S}
\end{center}
\end{figure}

In Fig. \ref{fig2S}, we show the values of ${\rm DI}_{ij}$, ${\rm CaMI}_{ij}$, 
${\rm CaMI}_{ji}$, and ${\rm MI}_{ij}$, for o neuron $i=1$ in a network
with $N=64$ neurons. There are conections from $i=1$ to $j=29,46,54$, and $65$.
In these case, we observe that ${\rm CaMI}_{ji} \approx {\rm MI}_{ij}$, therefore
${\rm TE}_{ij} \approx {\rm DI}_{ij} > 0$.
Moreover, there are conections from $j=3,5$, and $19$ to $i=1$, where 
${\rm DI}_{ij} < 0$. Finally, we can observe ${\rm DI}_{ij} \approx 0$ when there are
no connections.

\section*{Delay Analysis}

In Table 1, we show the case where no mapping delay is considered. We can
insert  a time delay by generating the symbol $s_2^n$ using the values of
$x_2^{n+\rm{delay}}$. In Fig. 1 we show the Directionality Index (DI) as a
function of mapping time step $\Delta t$ and delay in this mapping for $L=2$
and $L=6$. For the dynamical model used, we find that the DI is higher if $L=6$
than if $L=2$. In addition, the DI values are higher in the region where
$\Delta t<1.0$ and delay$<6$ms. This shows that a possible delay in
connections can be neglected in this case.

\begin{figure}[htbp!]
\begin{center}
\includegraphics[height=4cm,width=7cm]{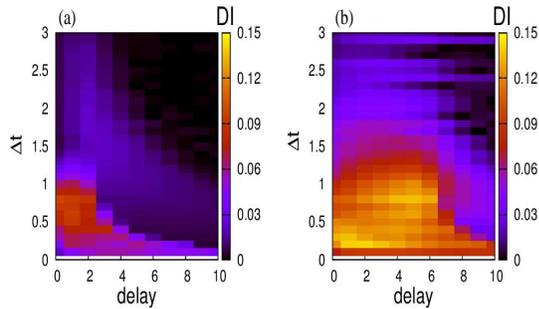}
\label{fig3S}
\caption{\small Directionality index ($DI$) as a function of mapping step time 
$\Delta t$ and delay in this mapping. (a) $L=2$ and (b) $L=6$.}
\end{center} 
\end{figure}

\section*{Additive noise}

\begin{figure}[htbp]
\begin{center}
\includegraphics[height=5.5cm,width=7cm]{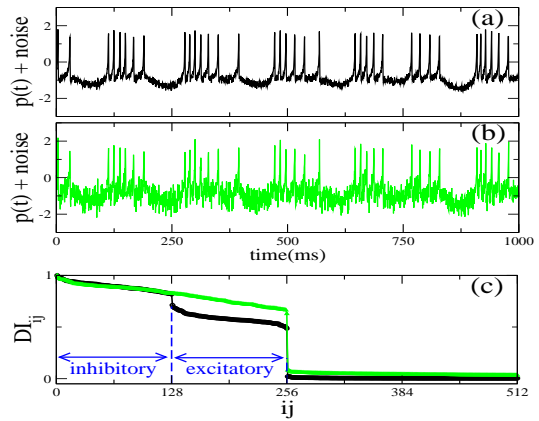}
\caption{Membrane potential for the neuron $i=32$ for additive noise standard 
deviation (a) $\sigma_a=0.1$ and (b) $\sigma_a=0.35$. (c) Normalised
Directionality index for $\sigma_a=0.1$ (black line) and $\sigma_a=0.35$ (green
line). We consider $\Delta t=0.5$, $L = 4$, and $g_c=0.1$.}
\label{fig4S}
\end{center}
\end{figure}

The additive noise is related to the imprecision of the equipment responsible 
for capturing the electrical signals in the neural membrane, so in our 
simulations we add to the values ​​of $p(t)$ a noise with zero mean and standard 
deviation $\sigma_a$. In Figs. \ref{fig4S}(a) and \ref{fig4S}(b) we observe
the change in the dynamics of the membrane potential of a network neuron under
the application of additive noise with $\sigma_a=0.1$ and $\sigma_a=0.35$,
respectively. The difference between the minimum and maximum values ​​reached by
the membrane potential of the HR model is approximately $3.5$, so
$\sigma_a=0.35$ corresponds to $10\%$ of this value. For $\sigma_a=0.1$ the
observed dynamics remains very similar to the case with no noise observed in
Fig. 1, however, when $\sigma_a=0.35$ the noise intensity can change the values
​of the symbolic sequence $S_i^L(n)$. In Fig. \ref{fig4S}(c) we see that the
DI calculation does not present significant changes when considering the
additive noise with $\sigma_a=0.1$ (black line). For $\sigma_a=0.35$ (green
line) it is no longer possible to distinguish excitatory connections from
inhibitory ones, but all $256$ connections are detected.

\section*{Information flow between Networks}

In many experimental cases it is not possible to directly measure the membrane 
potential of each neuron, but only an average field of a group of them, or a
brain region. Through the analysis of the mean field between two neural 
networks, we show that is possible to infer if distinct networks are connected
to each other, and identify the direction of the effective connectivity by the 
direction of the flow of information.

In order to do this analysis, we considered two random networks with $N=64$
neurons each, with average degree of intra connections within the networks
$K_{\rm intra}=24$ and average degree of inter connections between networks
$K_{\rm inter}=12$. To study the flow of information between the two networks, we
consider that there are only directed connections from neurons of network 1 to
neurons of network 2. In each of the networks we calculated the mean field of
the membrane potential and made the symbolic sequence using this time series.
The process of calculating $DI$ was performed in the same way as in the case of
isolated neurons.

\begin{figure}[htbp]
\begin{center}
\includegraphics[height=8cm,width=7cm]{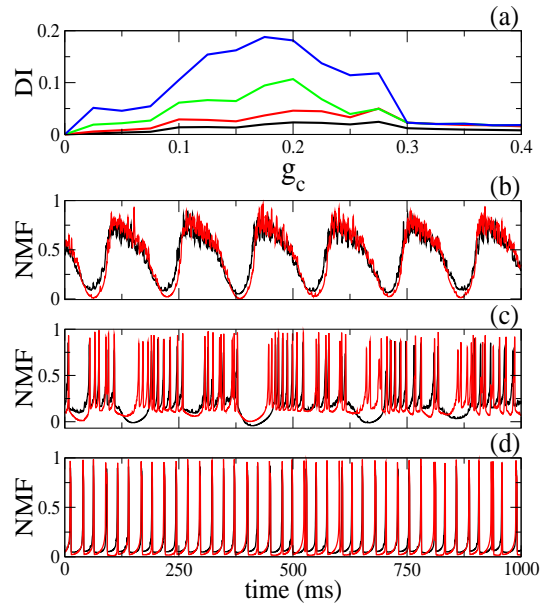}
\caption{(a) Directionality Index ($DI$) as a function of coupling strength
($g_c$). For $\Delta t= 0.25$ ms, where $L=1$ (black line), $L=2$ (red line),
$L=4$ (green line), and $L=8$ (blue line). Time evolution of the Normalised
Mean Field ($NMF$) for network 1 (black line) and network 2 (red line), both
with (b) $g_c=0.025$, (c) $g_c=0.175$, and (d) $g_c=0.275$.}
\label{fig5S}
\end{center}
 \end{figure}

In Fig. \ref{fig5S}(a) we show the values ​​obtained from the $DI$ as a function
of the intensity of the coupling $g_c$, where we set $\Delta t=0.25$ms and we
evaluate different sizes for the symbolic trajectory: $L=1$ (black line), $L=2$
(red line), $L=4$ (green line) and $L=8$ (blue line). We find that, as in the
case of two neurons, the highest $DI$ values ​​are observed when using symbolic
trajectories of size $L=8$. In this case, we observed that when the coupling is
low, the $DI$ values ​​are small, since the influence of the network dynamics 1
on the network 2 is smaller. For a coupling around $g_c=0.175$ we have the
highest calculated value of $DI$ and for $g_c>0.275$ the value of $DI$
decreases, tending to a constant value. This happens when the neurons of both
networks are roughly completely synchronous. The neurons had been completely
synchronous, thus the Transfer Entropy would be zero, resulting in a DI of
zero. To understand more about the dynamical behaviour leading to the curve
presented by the blue line in Fig. \ref{fig5S}(a), we analyse the temporal
evolution of the normalised mean field (NMF) for three values ​​of the coupling.
In Fig. \ref{fig5S}(b) we have $g_c=0.025$ and we observe that the NMF of
network 1 (black line) and network 2 (red line) show that the neurons of these
networks present the behaviour of bursting synchronisation, when neurons start
the bursting of firing activities roughly simultaneously. Firings are
asynchronous. In Fig. \ref{fig5S}(c) we have $g_c=0.175$ and the NMF of network
1 and 2 show that not only intra but also inter neurons are roughly
synchronous. Firing spikes in the NMF indicates intra synchronisation. Inter
synchronisation is evidenced by the fact that the curves are roughly identical.
These both factors are responsible for the high DI values. Finally, in Fig.
\ref{fig5S}(d) we have $g_c=0.275$ which is intense enough to make the
networks to almost fully synchronise.
 
Therefore, even in the case when we have only the data of the average field of 
networks, we show that it is possible to infer the effective directionality of
the connections in a similar way to the case between two neurons only. This 
method may be thus suitable to be considered for information flow studies in
different regions of the brain, analysing data obtained from several
experimental sources such as structural and functional MRI, diffusion tensor
imaging, magnetoencephalography, and electroencephalography.


\begin{thebibliography}{. . .}
\bibitem{barabasi04}
A.-L. Barab\'asi and Z. N. Oltvai, Nat. Rev. Genet. {\bf 5}, 101 (2004).
\bibitem{niemira04}
M. P. Niemira and T. L. Saaty, Int. J. Forecasting {20}, 573 (2004).
\bibitem{arenas08}
A. Arenas, A. D\'iaz-Guilera, J. Kurths, Y. Moreno, and C. Zhow, Phys. Rep.
{\bf 469}, 93 (2008).
\bibitem{bullmore09}
E. Bullmore and O. Sporns, Nat. Rev. Neurosci. {\bf 10}, 186 (2009).
\bibitem{gong09}
G. Gong, Y. He, L. Concha, C. Lebel, D. W. Gross, A. C. Evans, and C. Beaulieu,
Cerb. Cortex {\bf 19}, 524 (2009).  
\bibitem{ta10}
H. X. Ta, C. N. Yoon, L. Holm, and S. K. Han, BMC Syst. Biol. {\bf 4}, 70
(2010).
\bibitem{bastos16}
A. M. Bastos and J.-M. Schoffelen, Front. Syst. Neurosci. {\bf 9}, 1 (2016).
\bibitem{takigawa96}
M. Takigawa, G. Wang, H. Kawasaki, and H. Fukuzako, Int. J. Psychophysiol.
{\bf 21}, 65 (1996).  
\bibitem{baccala01}
A. Baccal\'a and K. Sameshima, Biol. Cybern. {\bf 84}, 463 (2001).  
\bibitem{cui15}
X.-M. Cui, W. S. Kim, D.-U. Hwang, and S. K. Han, Europhys. Lett. {\bf 110},
38001 (2015); Y. Chen, G. Rangarajan, J. Feng, and M. Ding, Phys. Lett. A
{324}, 26 (2004); N. Ancona, D. Marinazzo, and S. Stramaglia, Phys. Rev. E
{\bf 70}, 056221 (2004).
\bibitem{heuvel10}
M. P. van den Heuvel and H. E. H. Pol, Eur. Neuropsychopharmacol. {\bf 20},
519 (2010).
\bibitem{amblard11}
P.-O. Amblard and O. J. J. Michel, J. Comput. Neurosci. {\bf 30}, 7 (2011).
\bibitem{schreiber00}
T. Schreiber, Phys. Rev. Lett. {\bf 85}, 461 (2000).
\bibitem{runge15}  
J. Runge, Phys. Rev. E {\bf 92}, 062829 (2015).
\bibitem{sun15}
J. Sun, D. Taylor, and E. M. Bolt, SIAM J. Appl. Dyn. Syst. {\bf 14}, 73
(2015).
\bibitem{wang16}
W.-X. Wang, Y.-C. Lai, and C. Grebogi, Phys. Rep. {\bf 644}, 1 (2016).
\bibitem{rubido14}
N. Rubido, A. C. Mart\'i, E. Bianco-Mart\'inez, C. Grebogi, M. S. Baptista,
and C. Masoller, New J. Phys. {\bf 16}, 093010 (2014).
\bibitem{martinez16}
E. Bianco-Martinez, N. Rubido, C. G. Antonopoulos, and M. S. Baptista, Chaos
{\bf 26}, 043102 (2016).
\bibitem{massey90}
J. Massey, Causality, feedback and directed information (In Proc. Intl. Symp.
on Info. Theory and Its Applications, Waikiki, Hawai, USA, 27, 1990).
\bibitem{vlachos10}
I. Vlachos and D. Kugiumtzis, Phys. Rev. E {\bf 82}, 016207 (2010).
\bibitem{hindmarsh84}
J. L. Hindmarsh and R. M. Rose, Proc. R. Soc. London B {\bf 221}, 87 (1984).
\bibitem{pereda14}
A. Pereda, Nat. Rev. Neurosci. {\bf 15}, 250 (2014).
\bibitem{liu12}
Y. Liu and S. Aviyente, Comp. Math. Methods Med. {\bf 2012}, 635103 (2012).
\bibitem{bjorn14}
B. Schelter, M. Mader, W. Mader, L. Sommerlade, B. Platt, Y.-C. Lai, C.
Grebogi, and M. Thiel, Europhys. Lett. {\bf 105}, 30004 (2014).   
\bibitem{martinez15}
E. J. Bianco-Martinez, PhD Thesis, University of Aberdeen, 2015.
\bibitem{murilo16}
E. Bianco-Martinez and M. S. Baptista, arXiv:1612 05023v1.
\bibitem{barabasi99}
A.-L. Barab\'asi and R. Albert, Science {\bf 286}, 509 (1999).
\bibitem{timotheou10}
S. Timotheou, Comput. J. {\bf 53}, 251 (2010).
\bibitem{gelenbe89}
E. Gelenbe, Neural Comput. {\bf 1}, 502 (1989).
\bibitem{kaiser06}
M. Kaiser and C. C. Hilgetag, PLoS Comput. Biol. {\bf 2}, e95 (2006).
\bibitem{white86}
J. G. White, E. Southgate, J. N. Thomson, and S. Brenner, Phil. Trans. R. Soc.
Lond. B {\bf 314}, 1 (1986).  
\bibitem{hodgkin52}
A. L. Hodgkin and A. F. Huxley, J. Physiol. {\bf 117}, 500 (1952).
\bibitem{gyorgyi90}  
G. Gy\"orgyi, Phys. Rev. Lett {\bf 64}, 2957 (1990).
\bibitem{cao05}
X. J. Cao and D. Oertel, J. Neurophysiol. {\bf 94}, 821 (2005).
\bibitem{liu95}
K. S. Liu and P. W. Sternberg, Neuron {\bf 14}, 79 (1995).
\bibitem{varshney11}
L. R. Varshney, B. L. Chen, E. Paniagua, D. H. Hall, and D. B. Chklovskii,
PLoS Comput. Biol. {\bf 7}, e1001066 (2011).
\bibitem{gally03}
C. Gally, J. L. Bessereau JL, Med. Sci. {\bf 19}, 725 (2003).
\bibitem{celegans}
Connectome File Format-Datasets (Version 2.0). Available:
http://cmtk.org/viewer/datasets/. 
\end{thebibliography}
\end{document}